\documentclass[aps,prd,amsmath,amssymb,eqsecnum,nofootinbib,notitlepage,superscriptaddress]{revtex4}

\usepackage[retainorgcmds]{IEEEtrantools}
\usepackage{graphicx}
\usepackage{graphics}
\usepackage{color}
\usepackage{amsmath}

\usepackage{ulem}
\newtheorem{proposition}{Proposition}

\newtheorem{theorem}{Theorem}
\newtheorem{lemma}{Lemma}

\newcommand{\st}{\mathcal{M}}

\newcommand{\real}{\mathbb{R}}

\newcommand{\fab}[2]{\Gamma^{#1}_{{#2}'}}
\newcommand{\gtab}[2]{{\tilde{\Gamma}}^{#1}_{{#2}'}}
\newcommand{\eab}[2]{F^{{#1}'}_{{#2}'}}
\newcommand{\mab}{M^{a'}_{b'}}

\newcommand{\be}{\begin{equation}}
\newcommand{\ee}{\end{equation}}

\newcommand{\pd}[2]{\frac{\partial{#1}}{\partial{#2}}}
 
\newcommand{\godel}{{G\"{o}del}}

\newcommand{\ro}{\rho_0}
\newcommand{\lamb}{\bar{\lambda}}

\newcommand{\bi}{\begin{itemize}}
\newcommand{\ei}{\end{itemize}}

\begin{document}
\title{Motion of a gyroscope on a closed timelike curve}

\author{Brien C.\ Nolan}
\email{brien.nolan@dcu.ie}
\affiliation{Centre for Astrophysics \& Relativity, School of Mathematical Sciences, Dublin City
University, Glasnevin, Dublin 9, Ireland.}

\begin{abstract}
We consider the motion of a gyroscope on a closed timelike curve (CTC). A gyroscope is identified with a unit-length spacelike vector - a spin-vector - orthogonal to the tangent to the CTC, and satisfying the equations of Fermi-Walker transport along the curve. We investigate the consequences of the periodicity of the coefficients of the transport equations, which arise from the periodicty of the CTC, which is assumed to be piecewise $C^2$. 
We show that every CTC with period $T$ admits at least one $T-$periodic spin-vector. Further, either every other spin-vector is $T-$periodic, or no others are. It follows that gyroscopes carried by CTCs are either all $T-$periodic, or are generically not $T-$periodic. We consider examples of spacetimes admitting CTCs, and address the question of whether $T-$periodicity of gyroscopic motion occurs generically or only on a negligible set for these CTCs. We discuss these results from the perspective of principles of consistency in spacetimes admitting CTCs.    
\end{abstract}

\maketitle

\section{Introduction}

In this paper, we address the question of consistency in spacetimes admitting closed timelike curves (CTCs) \cite{friedman1990cauchy,echeverria1991billiard,novikov1992time,levanony2011extended,lewis1976paradoxes}. We consider this question from the perspective of the motion of a gyroscope carried by an observer moving on a CTC. 

In General Relativity (GR), time travel is associated with the presence of CTCs. This involves a point-particle approximation for the putative time machine. But there is a well-advanced theory of extended bodies in GR that builds on the concepts of the linear and angular momentum of the body, its multipole moments and its centre of mass \cite{harte2015motion,dixon1970dynamics1,dixon1970dynamics2,dixon1974dynamics3}. This theory provides an account of the motion of such bodies, including their self-interaction. Our ultimate aim is to ask if such bodies can undergo time travel, in the same way that point particles can (in the sense that CTCs exist in certain spacetimes). In this paper, we consider a simple example of extended structure that can be associated with a point particle - a gyroscope, which is modelled by a spin-vector, Fermi-Walker transported along the world-line of the particle (see e.g.\ Sections 6.5 and 40.7 of \cite{misner1973gravitation}). 

Of particular interest is the question of the consistency of such motion for extended bodies in general, and for gyroscopes in particular. The idea of self-consistency in time travel has been discussed from the perspective of both physics and philosophy  - perhaps most prominently in \cite{friedman1990cauchy,echeverria1991billiard,novikov1992time,levanony2011extended} and \cite{lewis1976paradoxes} respectively. Since a point particle has no internal structure or other distinguishing features, self-consistency is guaranteed for CTCs considered in isolation. But the same need not be the case for extended bodies.

It has been argued that CTCs need not be ruled out \textit{a priori}: what is relevant is the issue of self-consistency in spacetimes admitting CTCs 
\cite{friedman1990cauchy,echeverria1991billiard,novikov1992time,levanony2011extended}. The \textit{principle of self-consistency} (PSC) is elaborated in \cite{friedman1990cauchy}: ``the only solutions to the laws of physics that can occur locally in the real Universe are those which are globally self-consistent" (p.\ 1915). (This is sometimes referred to as the Novikov self-consistency principle; see endnote 10 of \cite{friedman1990cauchy}.) This principle is studied in some detail in these works, with perhaps surprising conclusions. In \cite{friedman1990cauchy}, it is shown that a class of 4-dimensional wormhole spacetimes admitting CTCs are \textit{benign} with respect to scalar field evolution. Roughly speaking, this means that if initial data for the scalar field, posed on a spacelike hypersurface $\Sigma$ and freely specified in a neighbourhood of a point $P\in\Sigma$, do not (already) lead to a self-consistent evolution, then an adjustment of the data outside this neighbourhood will lead to self-consistent evolution. Other spacetimes admitting CTCs are studied where scalar field evolution is not benign.

Self-interactions appear to cause problems for well-posed evolution, but not self-consistency. In their study of the classical motion of a billiard ball travelling backwards in time through a wormhole, and thereby (potentially) colliding with its earlier self, the authors of \cite{echeverria1991billiard} find large classes of initial trajectories that yield multiple (and in cases infinite) possible self-consistent evolutions. No cases are found where there are \textit{no} self-consistent evolutions. Novikov \cite{novikov1992time} also finds examples of self-consistent evolution of self-interacting mechanical systems in the presence of CTCs. 

Levanony and Ori study the propagation of rigid rods in Misner space-time \cite{levanony2011extended}. The original Misner space-time \cite{misner1967taub} is a region of 2-dimensional Minkowski spacetime, with identification of outgoing null hyperplanes generating CTCs. Adding (flat) Euclidean dimensions allows one to generate 3- and 4-dimensional Misner spacetimes. Levanony and Ori find a wide range of conditions which ensure consistency in the form of the absence of collisions of the rod with its (earlier) self. These conditions are phrased in terms of the length of the rod, and its velocity in the direction perpendicular to the spacetime direction along which the identification is made.

This perspective - that time-travel is not to be ruled out, but must not lead to inconsistencies - had previously been voiced by philosopher David Lewis \cite{lewis1976paradoxes}, who discusses self-consistency in an essay on ``The Paradoxes of Time Travel". In this essay, Lewis sets out his stall in the first sentence: ``Time travel, I maintain, is possible." (p.145). Lewis describes a clear conceptual framework for time travel. This is done in philosophical rather than mathematical language, but his framework would be familiar to relativists, and, moreover, this framework captures essential features of how relativists describe time travel\footnote{One feature likely to make relativisits uncomfortable is Lewis's use of ``external time" (\cite{lewis1976paradoxes}, p.146), which sounds suspiciously like universal Newtonian time or indeed a global time coordinate. The existence of a global time coordinate implies that spacetime is stably causal (see e.g.\ Theorem 8.2.2 of \cite{wald1984general}), and hence obeys chronology - i.e.\ is devoid of CTCs \cite{minguzzi2019lorentzian}. However this can be remedied by associating external time with time measured by a fixed reference observer on an open, timelike curve. With this interpretation, external time need not be identified with a Newtonian concept of universal time. If we accept this view, then Lewis's use of external time does not seriously undermine the compatibility of his framework with the relativistic perspective. Lewis's work has strongly influenced philosophical discussions of time travel and causality, including in characterizing time travel in terms of a discrepancy between external time and the ``personal time" of the time traveller (\cite{lewis1976paradoxes}, p.146). See \cite{wasserman2017paradoxes}.}. He goes on to discuss the grandfather paradox, whereby a time traveller - Tim - travels back in time to decades before his birth to kill his grandfather. Having considered the alternative, Lewis's conclusion is unequivocal: ``So Tim cannot kill Grandfather" (p.150). His argument is essentially that events must be compatible with all other relevant facts. The key fact of relevance to this imagined scenario - where Tim has travelled to 1920, bought a rifle, spent time training himself in its use and tracked down Grandfather in 1921 - is Tim's existence. His attempted killing of Grandfather is impossible, not being compatible with this fact. 

Lewis's central point is that the possibility of time travel should be admitted, provided it does not lead to inconsistencies in the shape of the occurrence of two events which are \textit{not} compatible.

In this paper, we study a simple non-interacting system: a gyroscope carried by a closed timelike curve. In this system, we see an opportunity to investigate what the laws of physics say about CTCs, and in particular, about consistency in the presence of CTCs. The relevant laws of physics in this context are encapsulated in the following: \textit{Gyroscopes exist. Given a timelike curve $\gamma$ representing the history of an observer in spacetime, a gyroscope carried by this observer is identified with a spin-vector. This is a unit length spacelike vector $s^a$ that is orthogonal to $u^a$, the tangent to $\gamma$, and that satisfies the equations of Fermi-Walker transport along $\gamma$}. 

Gyroscopes precess. This presents difficulties for consistent evolution along CTCs. At proper time $t_0$, the gyroscope - or spin-vector $s^a$ - has a certain orientation, $s^a(t_0)\in T_{\gamma(t_0)}(\st)$. At a later time $t_1>t_0$, it has another orientation, $s^a(t_1)\in T_{\gamma(t_1)}(\st)$. If $\gamma$ is closed, with $\gamma(t_0)=\gamma(t_1)$, we may well find that due to precession, $s^a(t_0)\neq s^a(t_1)$. This presents an inconsistency that is not present at the level of the CTC itself, thought of as the history of a structureless, featureless point particle. For consistent evolution, we would require that $s^a(t_0)= s^a(t_1)$. Note that this equation makes sense as both vectors lie in the same tangent space $T_{\gamma(t_0)}(\st)=T_{\gamma(t_1)}(\st)$.   

Our aim in this paper is to determine the extent to which the motion of a gyroscope on a CTC is consistent. Our conclusion is that it almost never is. Classical physics intervenes, and rules out consistent histories in spacetimes with CTCs. 

Our main results are contained in the following section, where we consider the space of solutions of the equations of Fermi-Walker (FW) transport along a closed timelike curve $\gamma$. The angle and volume preserving nature of FW transport implies that it induces a rotation on the space of spin-vectors as we complete one loop of the CTC. We can deduce from this the possibilities for consistent evolution of the spin-vector: of the three independent spin-vectors carried by an observer on a CTC, one is always periodic (i.e.\ returns to its original orientation on completion of one loop of the CTC). Of the other two, they are either both also periodic (consistent motion of the gyroscope), or neither is periodic (inconsistent motion of the gyroscope).  The closed nature of $\gamma$ implies that the FW transport equations - a system of linear ODEs - have a coefficient matrix whose entries are piecewise continuous periodic functions. This allows us to draw some general conclusions about the structure of solutions of the FW transport equations by applying the Floquet-Lyapunov theorem \cite{yakub1975, magnus2013hill}. In Section 3, we study some examples of gyroscopic motion in spacetimes admitting CTCs. These serve to illustrate the results of Section 2, and hint at the general picture. We make some conclusions in Section 4, where we argue that consistent evolution is generically not possible. We follow the conventions of Wald \cite{wald1984general}.

\section{Equations of motion and their solutions} 
In this section, we study the equations of motion of a gyroscope along a closed timelike curve $\gamma:[a,b]\to M:t\in\mathbb{R}\to\gamma(t)\in M$, with $\gamma(a)=\gamma(b)$. We parametrize by proper time $t$. We immediately replace $\gamma$ with its periodic extension (also called $\gamma$), defined for all $t\in\real$, so that in any local coordinate system in which $\gamma$ given is by $t\mapsto x^a(t)$, we have 
\be x^a(t+T) = x^a(t)\quad \hbox{for all } t\in\mathbb{R}. \label{ctc-period}
\ee
We refer to any such quantity as being $T-$periodic, and we note that $T=b-a$ (so $t=b$ is the first time $t>a$ at which the curve meets itself). We assume that $\gamma$ is piecewise $C^2$. 
Then there is a countable set of values $\mathbb{D}$, which has finite intersection with any closed interval, such that the velocity $u^a$ and acceleration $a^a$ of $\gamma$ are defined for all $t\in\mathbb{R}$ and are continuous on $\mathbb{R}\setminus\mathbb{D}$. 
We assume also that the velocity and acceleration have left- and right-hand limits at points of $\mathbb{D}$. It follows that the velocity and acceleration are likewise $T-$periodic (and are piecewise $C^1$ and piecewise $C^0$ respectively):
\be u^a(t+T)=u^a(t),\quad a^a(t+T)=a^a(t)\quad \hbox{for all } t\in\mathbb{R} \label{vel-acc-periodic}. \ee
It may be worth repeating that (\ref{vel-acc-periodic}), and similar equations below, have the apparently troubling feature of comparing tensors at different spacetime events. However periodicity of $\gamma$ means that this is not the case: $u^a(t)$ and $u^a(t+T)$ both lie in the tangent space $T_{\gamma(t)}(\st)=T_{\gamma(t+T)}(\st)$.
As we have parametrized by proper time, we have the usual relations 
\be g_{ab}u^au^b=-1,\quad g_{ab}u^aa^b=0. \label{ips-1} \ee

We identify a gyroscope carried by an observer moving on $\gamma$ with a \textit{spin-vector} $s^a$ along $\gamma$. This satisfies the Fermi-Walker transport equations along $\gamma$: 
\be u^c\nabla_c{s^a}=(u^aa_b-a^au_b)s^b. \label{FW-eqn} \ee
Fermi-Walker (FW) transport along an arbitrary timelike curve generalises the concept of parallel transport along a geodesic. A key property is that inner products are preserved by FW transport: if $s^a$ and $t^a$ are both solutions of (\ref{FW-eqn}), then $g_{ab}s^at^b$ is constant along $\gamma$. 
It follows that norms of FW-transported vectors are also constant along $\gamma$. Note also that $u^a$ itself is FW-transported along $\gamma$: taking $s^a=u^a$ in (\ref{FW-eqn}), we see that both sides evaluate to $a^a$.  
We define a spin-vector along $\gamma$ to be an FW-transported vector that is orthogonal to $u^a$ along $\gamma$. As inner products are conserved, we see that this condition needs to be applied only at some initial time. Since norms are conserved, we can also assume without loss of generality that spin-vectors have unit length. 

Let us now fix an initial point $z^a=x^a(0)$ on $\gamma$; we use $z=\gamma(0)$ to label the corresponding space-time event. Our aim is to determine if a spin-vector (gyroscope) moving on the CTC maintains its orientation when it completes one circuit of the CTC, starting at $\gamma(0)=z$ and terminating at $\gamma(T)=z$. In other words, we wish to determine if spin-vectors along $\gamma$ are $T-$periodic\footnote{A spin-vector cannot have period less than $T$. With reference to the CTC $\gamma$, we use the phrase $T-$periodic in the usual sense whereby $T$ is the minimal positive value of $t$ for which $x^a(t+T)=x^a(t)$ for all $t\in\mathbb{R}$. So for $t<T$, $s^a(t)\notin T_z(\st)$, and so we cannot possibly have $s^a(t)=s^a(0)$.}. We wish to consider all possible spin-vectors to determine if $T-$periodicity is a generic feature of gyroscopic motion along a CTC. Thus we wish to determine global properties of the general solution of (\ref{FW-eqn}). 

These properties are captured by the \textit{Fermi-Walker propagator} (FW propagator) of (\ref{FW-eqn}). This is a 2-point tensor $\fab{a}{b}(x(t),z)$  along $\gamma$. We use unprimed indices at $x(t)=x^{a}(t)$ and primed indices at the initial point $z=x(0)$. The FW propagator satisfies the transport equation 
\be u^c\nabla_c\fab{a}{b} = (u^aa_b-a^au_b)\fab{b}{b},\quad t\geq0 \label{FW-prop} \ee
and the initial condition 
\be [\fab{a}{b}] := \fab{a}{b}(0) = \delta^{a'}_{b'}.
\label{FW-prop-ic} \ee
The word `propagator' is used because $\fab{a}{b}$ propagates initial data to generate solutions of (\ref{FW-eqn}): the unique solution of the initial value problem comprising (\ref{FW-eqn}) and the initial condition 
\be s^b(0) = s_0^{b'} \in T_{z}(\st) \label{ics} \ee
can be written as  
\be s^a(t) =\fab{a}{b}(x(t),z) s_0^{b'},\quad t\geq 0.\label{FW-prop-data}\ee

Thus $\fab{a}{b}$ is a fundamental (or transition) matrix of the linear ODE (\ref{FW-eqn}) \cite{coddington1955theory}. The coefficient matrix 
\be C^a_b := u^aa_b-a^au_b \label{coeff-def} \ee
is periodic and piecewise continuous, and so we can apply results of Floquet theory \cite{yakub1975, magnus2013hill}. We restate some key results, which require minor modifications to take account of the geometric setting. 

The FW propagator defines a continuous family of volume elements at $z=x(0)$ defined by  
\be \Delta_{a'b'c'd'}(x(t),z) = \epsilon_{abcd}\fab{a}{a}\fab{b}{b}\fab{c}{c}\fab{d}{d}. \label{eq:vol-el}
\ee
This quantity is a scalar at $x(t)$, and a 4-form (volume element) at $z=x(0)$. Taking the derivative and using (\ref{FW-prop}) yields the following:

\begin{lemma}{(Liouville's formula.)}
\be \Delta_{a'b'c'd'}(x(t),z) = \Delta_{a'b'c'd'}(x(0),z)\exp\{\int_0^t C^a_a(t') dt'\}, 
\label{Liouville}
\ee
and since $C^a_a=0$ and $[\fab{a}{b}]=\delta^{a'}_{b'}$,
\be \Delta_{a'b'c'd'}(x(t),z) = \epsilon_{a'b'c'd'}(z). \label{D-is-epsilon}
\ee
Thus the volume element (\ref{eq:vol-el}) is constant. \hfill$\blacksquare$
\end{lemma}
This expresses the norm-preserving character of FW transport.

Let us write $\fab{a}{b}(x(t),z)=\fab{a}{b}(t)$. Then the \textit{monodromy matrix} \cite{yakub1975} 
\be \mab:=\fab{a}{b}(T) \label{mono-def} \ee 
plays a particularly important role in our analysis. Since $T$ is the period of the CTC, so that $\gamma(T)=\gamma(0)=z$, the monodromy matrix generates (via (\ref{FW-prop-data})) a linear transformation 
\be M:T_z(\st)\to T_z(\st): s^{a'}_0\mapsto s^{a'}(T)=\mab s^{b'}_0. \label{mono-map}\ee
By (\ref{D-is-epsilon}), the determinant of this mapping is 1. The velocity vector $u^a$ is a $T-$periodic solution of the FW transport equations (\ref{FW-eqn}), and so yields an eigenvector of $M$ with eigenvalue 1: 
\be  u^{a'}(0)=u^{a'}(T) = \mab u^{b'}(0). \label{v-eval} \ee
The first equality here arises from $T-$periodicity of $u^a$, and the second is an application of (\ref{FW-prop-data}). As noted previously, an FW transported vector that is initially orthogonal to $u^a$ remains orthogonal to $u^a$ along $\gamma$, and has constant norm. Hence $M$ also induces a linear transformation 
\be M^{\perp}: T_z^{\perp}(\st)\to T_z^{\perp}(\st): s^{a'}_0\mapsto s^{a'}(T)=\mab s^{b'}_0, \label{mono-map-perp}\ee
where for $x\in\st$ on $\gamma$,
\be T_x^{\perp}(\st) = \{v^a\in T_x(\st): g_{ab}u^av^b=0\}. \label{T-perp-def} \ee
The transformation $M^{\perp}$ preserves inner products (and hence orientation) and norms, and so can be identified with an element of $SO(3)$ acting on 
\be \mathbb{S}^2\simeq T_x^{\perp,1}(\st) = \{v^a\in T_x(\st): g_{ab}u^av^b=0, g_{ab}v^av^b=1\}. \label{T-perp-unit-def} \ee
It follows that $M^{\perp}$ has eigenvalues $\{1,e^{i\theta},e^{-i\theta}\}$ for some $\theta\in\mathbb{R}$. (In the case $\theta=0$, $M^{\perp}$ is the identity mapping and we must have $\mab=\delta^{a'}_{b'}$.)

Let $s^{a'}_1$ be a (unit) eigenvector of $M^{\perp}$ with eigenvalue equal to 1, and consider the solution of the FW transport equations (\ref{FW-eqn}) with initial value 
\be s^{a'}(0) = s^{a'}_1. \label{fixed-initial} \ee
Propagating this initial value using (\ref{FW-prop-data})and (\ref{mono-def}), we find 
\be s^{a'}(T) = \mab s^{b'}_1 = s^{a'}_1 = s^{a'}(0). 
\label{periodic-spin}
\ee
That is, the spin-vector with initial value $s^{a'}_1$ is $T-$periodic. This follows by uniqueness of solutions of (\ref{FW-eqn}): the solution for $s^{a'}(t)$ on $[0,T]$ is repeated on $[nT,(n+1)T]$, $n\in \mathbb{N}$, so that $s^{a'}(t+T)=s^{a'}(t)$ for all $t\in\mathbb{R}$. 

Thus we can write down the following result:

\begin{proposition}\label{prop2} Every $T-$periodic closed time-like curve admits a $T-$periodic spin-vector. \hfill$\blacksquare$
\end{proposition}


The remaining eigenvalues of $M^\perp$ are, in general, complex, and are associated with a pair of unit spacelike vectors, orthogonal to both $u^{a'}(0)$ and $s^{a'}_1$. We can take these to be $v^{a'}_1, w^{a'}_1 \in T^{\perp,1}_z(\st)$ with
\begin{eqnarray} \mab v^{b'}_1 &=& \cos\theta v^{a'}_1 - \sin\theta w^{a'}_1, \label{u-evec} \\ 
\mab w^{b'}_1 &=& \sin\theta v^{a'}_1 +\cos\theta w^{a'}_1 \label{w-evec} 
\end{eqnarray}
for some $\theta\in\real$. The associated eigenvectors are $v^{a'}_1\pm i w^{a'}_1$ with eigenvalues $e^{\pm i \theta}$ respectively. Then $\{s^{a'}_1,v^{a'}_1,w^{a'}_1\}$ forms an orthonormal basis for $T_z^{\perp,1}(\st)$, which we recognize as being the set of initial data for spin-vectors along the $T-$periodic CTC $\gamma$. We can make the identification of $T_z^{\perp,1}(\st)$ with $\mathbb{S}^2$ explicit by writing $s^{a'}_0\in T_z^{\perp,1}(\st)$ as 
\be s^{a'}_0 = a s^{a'}_1+b v^{a'}_1 + c w^{a'}_1,\quad (a,b,c)\in\mathbb{S}^2\subset \mathbb{R}^3. \label{s0-s2} 
\ee
At time $t=T$, the solution of (\ref{FW-prop}) with initial data $s^{a'}(0)=s^{a'}_0$ is given by $\mab s^{b'}_0$ so that   \be s^{a'}(T) = as^{a'}_1 +(b\cos\theta + c\sin\theta)v^{a'}_1+(c\cos\theta - b\sin\theta)w^{a'}_1. 
\label{gen-sol-T} 
\ee
This leads to the following result:

\begin{proposition}\label{prop3}
Let $\gamma$ be a $T-$periodic CTC. Then either 
\bi
\item[(i)] every spin-vector along $\gamma$ is $T-$periodic, or \item[(ii)] in the set of initial data for spin-vectors along $\gamma$,  initial data which yield a $T-$periodic spin-vector along $\gamma$ form a set of measure zero. 
\ei
\end{proposition}

\noindent\textbf{Proof:} In (\ref{gen-sol-T}), we can assume that $\theta\in [0,2\pi)$. If $\theta=0$, then $s^{a'}(T)=s^{a'}(0)$ for all $(a,b,c)\in\mathbb{S}^2$ and (i) holds. If $\theta\neq 0$, then $s^{a'}(T)=s^{a'}(0)$ if and only if $(a,b,c)=(\pm1,0,0)\in\mathbb{S}^2$. Thus there are just two points of $\mathbb{S}^2$ corresponding to a $T-$periodic spin vector - a set of measure zero - and (ii) holds. \hfill$\blacksquare$

\subsection{The Floquet-Lyapunov Theorem}

In the following section, we consider examples of CTCs in different spacetimes and determine which of the two outcomes of Proposition \ref{prop3} hold. Ideally, we would be able to make a statement about the generic behaviour of spin-vectors along CTCs. Of course there may not be any such statement: there might not be a generic outcome vis a vis Proposition \ref{prop3} in the sense of a conclusion that applies to all but a set of measure zero in the set of CTCs. Drawing such a conclusion would require determining the eigenvalues of the monodromy matrix $\mab$. The setting is a first order linear system with periodic coefficients \cite{yakub1975, magnus2013hill}. Such equations display a wide variety of behaviours, but the Floquet-Lyapunov theorem provides a degree of order. We state this in reference to a slight generalisation of the FW transport equation (\ref{FW-eqn}) and note that relative to the classical form of this result, some geometric dressing is required. See \cite{yakub1975} or \cite{coddington1955theory} for a proof of the corresponding result for systems of the form $x'(t)=A(t)x(t)$ where $x(t)\in \mathbb{C}^n$ and $A$ is a $T-$periodic $n\times n$ matrix. These proofs extend more or less directly to the geometric setting, but we provide a proof in the appendix to clarify its geometric content.


\begin{theorem}\label{FL-theorem}{(Geometric Floquet-Lyapunov theorem \cite{yakub1975}.)} Let $\fab{a}{b}(t)$ be a propagator for the linear transport equation along $\gamma$ given by
\be v^b\nabla_b s^a = C^a_b(t) s^b \label{prop-gen} \ee
where $C^a_b(t)$ is $T-$periodic and piecewise $C^0$ along $\gamma$, so that $\fab{a}{b}$ satisfies the equations 
\be v^b\nabla_b \fab{a}{b} = C^a_b(t) \fab{b}{b},\quad [\fab{a}{b}] = \delta^{a'}_{b'}. \label{prop-gen-matrix} 
\ee
Then $\fab{a}{b}$ may be written 
\be \fab{a}{b}(t) = P^a_{c'}(t)E^{c'}_{b'}(t), \label{F-L-Theorem}
\ee
where $P^a_{c'}(t)\equiv P^a_{c'}(x(t),z)$ is a $T-$periodic 2-point tensor along $\gamma$ with $[P^a_{c'}]=\delta^{a'}_{c'}$, and $E^{c'}_{b'}(t)$ is an exponential along $\gamma$ - that is, $E^{c'}_{b'}(t)$ is a 2-point tensor (scalar at $x(t)$ on $\gamma$, type (1,1) at $z=x(0)$) that solves an initial value problem  of the form 
\be \frac{dE^{a'}_{b'}}{dt} = A^{a'}_{c'}E^{c'}_{b'},\quad E^{a'}_{b'}(0) =\delta^{a'}_{b'}\label{exp-def-de}\ee
where $A^{a'}_{b'}$ is a non-singular tensor of type (1,1) at $z$. \hfill$\blacksquare$
\end{theorem}

\section{Examples}

The geometric Floquet-Lyapunov theorem applies to (\ref{FW-eqn}), as $C^a_b=v^aa_b-a^av_b$ satisfies the required periodicity and continuity hypotheses. In terms of (\ref{F-L-Theorem}), it follows that the monodromy matrix is given by $\mab=E^{a'}_{b'}(T)$. While this object is in some sense relatively simple, being an exponential tensor (and so essential being the exponential of a constant matrix), the connection between the coefficients $A^{a'}_{b'}$ of (\ref{exp-def-de}) and $C^a_b$ of (\ref{prop-gen}) is highly non-trivial \cite{magnus2013hill}. Nevertheless, we will attempt some general conclusions in the following section. Here, to illustrate possibilities, we consider gyroscopic motion along CTCs in a number of different spacetimes. 

\subsection{Stationary cylindrically symmetric spacetimes}

A number of examples of CTCs arise in stationary, cylindrically symmetric spacetimes. Included here are the historically important examples of \godel's spacetime \cite{godel1949example} and van Stockum's spacetime \cite{van1937gravitational}, which forms the basis of a Tipler machine \cite{tipler1974rotating}. For this reason, we will review the relevant equations in this general setting, and specialise to draw conclusions about specific examples. We note that the motion of a gyroscope along geodesics (rather than CTCs) of \godel's spacetime has been considered in \cite{bini2019agodel} and \cite{bini2019bgodel}.

We can write the line element of a stationary, cylindrically symmetric spacetime as 
\be ds^2 = -Fd\tau^2+2Md\tau d\phi + Ld\phi^2 + H_1dr^2+H_2d\zeta^2, \label{stn-cyl-lel} \ee
where the metric functions $F,M,L,H_1$ and $H_2$ depend only on $r$, and $F,H_1,H_2$ are positive. We also have the restriction $FL+M^2>0$ to ensure that the signature is Lorentzian. We use coordinates $(\tau,r,\phi,z)$ that are closely related but not identical to the Weyl-Papapetrou coordinates (see e.g.\ \cite{tipler1974rotating} and section 13.1 of \cite{griffiths2009exact}). This line element admits (at least) three Killing vector fields: $\pd{}{\tau}$, which is timelike and so yields stationarity, and the axial and translational Killing vector fields $\pd{}{\phi}$ and $\pd{}{\zeta}$ respectively, generating the cylindrical group of motions. The axis corresponds to the 2-dimensional submanifold $\{r=0\}$, and we can assume that $\zeta$ ranges over the whole real line. The azimuthal coordinate $\phi\in[0,2\pi)$ is subject the the usual $2\pi$-periodic identification. 

The metric admits a 3-parameter family of CTCs \be \gamma:t\in\mathbb{R}\mapsto (\tau_0,r_0,\phi(t),\zeta_0)\in\st \label{stn-cyl-pctc} \ee where $\tau_0, r_0$ and $\zeta_0$ are constants. For convenience, we'll refer to such a curve as a \textit{circular CTC with radius $r_0$}.  More accurately, these curves are always closed and periodic, and are timelike provided 
\be L(r_0)<0. \label{CTC-stn-cyl} \ee
As above, we take $t$ to be proper time along each curve. The tangent to $\gamma$ is 
\be u^a = \phi'(t)\left(\pd{}{\phi}\right)^a, \label{ctc-tgt} \ee
and so 
\be \left(\frac{d\phi}{dt}\right)^2=-\frac{1}{L(r_0)}. \label{phi-dot} \ee 
Then we can write down the (proper time) period of these CTCs:
\be T_{\gamma}=2\pi\sqrt{-L(r_0)}. \label{period} \ee

It is straightforward to write down the Fermi-Walker transport equations describing the evolution of a spin-vector $s^a$ along $\gamma$. Recall that we also impose the orthogonality condition 
\be g_{ab}u^as^b=0. \label{ortho} \ee 

We find that 
\be s_{(1)}^a(t) = S_1\left(\pd{}{\zeta}\right)^a,\quad S_1=\hbox{ constant } \label{spin1-stn-cyl} \ee
is always a spin-vector along $\gamma$. Furthermore, this spin-vector clearly satisfies 
\be s_{(1)}^a(t+T_\gamma) = s_{(1)}^a(t)\quad \hbox{for all } t\in \mathbb{R}, \label{s1-periodic} \ee
yielding a $T_\gamma-$periodic spin-vector along $\gamma$. 
Spin-vectors orthogonal to $s_{(1)}^a$ may be written as 
\be s^a = S_2(t)\left(\pd{}{\tau}\right)^a + S_3(t)\left(\pd{}{\phi}\right)^a + S_4(t)\left(\pd{}{r}\right)^a. \label{spin-cyl-others} \ee
Imposing (\ref{ortho}) and (\ref{FW-eqn}) yields a first order system in $\real^3$ with a zero order constraint, linear in the $S_i, i=2,3,4$ (and so with two degrees of freedom). If the non-negative quantity $\Omega(r_0)$ vanishes (see (\ref{cyl-omega}) below), then the $S_i$ are constant along $\gamma$, and all corresponding spin-vectors are periodic. If $\Omega(r_0)>0$, these equations reduce to a single second order equation in $S(t)\in\{S_2,S_3,S_4\}$: all three variables satisfy this equation. This master equation reads 
\be S''(t) + \Omega^2(r_0)S(t) = 0,\label{cyl-master} \ee
where 
\be \Omega^2(r_0) = \left.\frac{(ML'-LM')^2}{4H_1L^2(FL+M^2)}\right|_{r=r_0}.\label{cyl-omega} 
\ee
This quantity is non-negative - this follows from the conditions above on the metric functions. We need only consider the case where it is positive. Considering $S$ as a function $S:t\in\mathbb{R}\mapsto S(t)\in\mathbb{R}$, we see that every (non-trivial) solution $S$ of (\ref{cyl-master}) has period 
\be T_s = \frac{2\pi}{\Omega(r_0)}. \label{spin=period} \ee
Thus the corresponding spin-vector is $T_\gamma-$periodic if and only if $T_\gamma$ is an integer multiple of $T_s$. We can summarise as follows: 

\begin{proposition}\label{stn-cyl-prop}
Every spin-vector carried by a circular CTC $\gamma$ with radius $r$ is $T_\gamma-$periodic if and only if 
\be \lambda(r):=\frac{(ML'-LM')^2}{4H_1|L|(FL+M^2)} = n^2 \quad\hbox{for some } n\in \mathbb{N}. \label{spin-period-cyl}\ee
If this condition is not met, then there is exactly one spin-vector along $\gamma$ which is $T_\gamma-$periodic. \hfill$\blacksquare$
\end{proposition}

We now apply this proposition to a selection of stationary, cylindrically symmetric spacetimes admitting circular CTCs. 

\subsubsection{\godel's universe}
The discovery of \godel's solution of the Einstein equations \cite{godel1949example} occupies an important place in General Relativity, both scientifically and historically \cite{ellis2000editor,ozsvath2003godel,rindler2009godel}. In particular, this was the first spacetime in which CTCs were identified. There is a large body of literature on this spacetime, which we will not review here (see \cite{ellis2000editor} and the more recent \cite{nolan2020causality} for references). The line element of \godel's universe can be written in the form (\ref{stn-cyl-lel}) with 
$F=H_1=H_2=1$ and 
\be L(r)=\alpha^{-2}(\sinh^2\alpha r-\sinh^4\alpha r),\quad M(r) = -\sqrt{2}\alpha^{-1}\sinh^2\alpha r. \label{godel-metric-fns}\ee
The real, positive parameter $\alpha$ sets an overall scale in this homogeneous spacetime: the Einstein-perfect fluid field equations give the energy density $\rho$ and pressure $P$ as 
\begin{eqnarray} 8\pi\rho &=& 2\alpha^2-\Lambda,\label{godel-density}\\
8\pi P &=& 2\alpha^2+\Lambda,\label{godel-pressure}
\end{eqnarray}
where $\Lambda$ is the cosmological constant. 

\begin{figure}[ht!]
\begin{center}
\includegraphics[width=10cm]{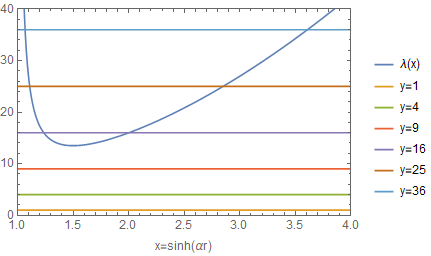}
  \end{center}
\caption{Plot of the function $\lambda(r)$ of Proposition 
\ref{stn-cyl-prop} for \godel's spacetime. The graph intersects each horizontal line $y=n^2, n\geq4$ twice; the value of $r$ at each intersection corresponds to a circular CTC along which all spin-vectors are periodic with the same period as the CTC on which they travel. Gyroscopic motion is consistent on these circular CTCs. As noted in the text, we say that the circular CTCs of this spacetime demonstrate the \godel\ profile with (in this case) $n=4$.}
\label{fig:1}
\end{figure}

From (\ref{godel-metric-fns}), we see that there are circular CTCs with radius  $r$ whenever 
\be \sinh \alpha r >1. \label{godel-ctc-con} \ee
A straightforward calculation yields 
\be \lambda(r) = \frac{2\sinh^6\alpha r}{\sinh^2\alpha r - 1}. \label{godel-lambda} \ee
We note that the singular behaviour at $\alpha r = \ln(1+\sqrt{2})$ is due to the fact that the curve (\ref{stn-cyl-pctc}) becomes null at this value of $r$. 
For $\rho=\sinh\alpha r>1$, $\lambda(r)$ in (\ref{godel-lambda}) approaches $+\infty$ asymptotically as $\rho\to 1^+$ and as $\rho\to+\infty$, and has a global minimum of $27/2\in(3^2,4^2)$. So we can summarise as follows (see Figure \ref{fig:1}): 

\begin{proposition}\label{Prop-Godel} For each $n\geq4$, there are two circular CTCs $\gamma$ in \godel's universe along which all spin-vectors are $T_\gamma-$periodic. For each $n\geq4$, these correspond to the two values of $r$ which solve $\lambda(r)=n^2$. Along all other circular CTCs, there is exactly one $T_\gamma-$periodic spin-vector. \hfill$\blacksquare$
\end{proposition}

A nearly identical conclusion will hold in some of the other examples below, so for convenience we will refer to the outcome described in this proposition by saying that the \textit{\godel\ profile with $n= n_0$} applies, where $n_0=4$ in this case. 

\subsubsection{Som-Raychaudhuri spacetime}

The Som-Raychaudhuri spacetime  is a stationary, cylindrically symmetric spacetime filled with charged dust \cite{som1968cylindrically}. Its causal properties have been investigated in \cite{paiva1987time}: like \godel's universe, this homogenous spacetime is totally vicious (meaning that CTCs pass through every event), but does not contain any closed causal geodesics. The line element is given by (\ref{stn-cyl-lel}) with $F(r)=1$, 
\be L(r)=r^2(1-\alpha^2r^2),\quad M(r) =\alpha r^2 \label{LM-SR} \ee
and
\be H_1(r)=H_2(r) = e^{(\beta^2-\alpha^2)r^2}.\label{H-SR}\ee
The positive constants $\alpha,\beta$ determine the energy and charge densities of the fluid which are given by (respectively)
\be 4\pi\rho = (2\alpha^2-\beta^2)e^{(\alpha^2-\beta^2)r^2},\quad 4\pi\sigma=2\alpha\beta e^{(\alpha^2-\beta^2)r^2}. \label{SR-matter}\ee
We will assume the energy condition $2\alpha^2-\beta^2>0$. 

\begin{figure}[ht!]
\begin{center}
\includegraphics[width=10cm]{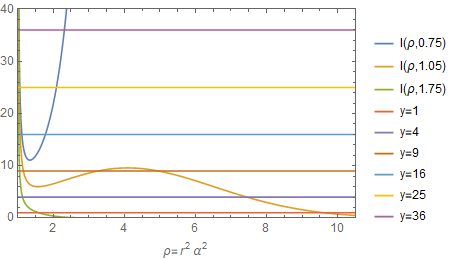}
  \end{center}
\caption{Plot of the function $\lambda(r)=l(\rho,b)$ of Proposition 
\ref{stn-cyl-prop} for Som-Raychaudhuri spacetime. The dark blue graph has $b=0.75$ and exemplifies part (i) of Proposition \ref{Prop-SR}; circular CTCs follow the \godel\ profile with $n=4$. The orange graph has $b=1.05$ and so exemplifies part (ii) of the proposition. There are three circular CTCs corresponding to $n=3$, and just one corresponding to all other values of $n\geq1$. The green graph has $b=1.75$, and exemplifies part (iii) of the proposition. Here, there is a unique circular CTC corresponding to each value of $n$. Recall that the intersections of the graph of $l(\rho,b)$ and the horizontal lines $y=n^2$ correspond to CTCs for which all gyroscopes are $T_\gamma$-periodic (and so consistent). All other circular CTCs have generically inconsistent gyroscopes.}
\label{fig:2}
\end{figure}

Circular CTCs occur in the region $\alpha r>1$. We calculate

\be \lambda(r) = \frac{\alpha^6r^6e^{(\alpha^2-\beta^2) r^2}}{\alpha^2r^2-1}. \label{SR-lambda} \ee
We can write this as 
\be \lambda(r) = l(\rho,b) := \frac{\rho^3e^{(1-b)\rho^2}}{\rho-1},\quad \rho=\alpha^2 r^2,\quad b = \beta^2/\alpha^2 \in [0,2). \label{SR-new-var} \ee
Then circular CTCs occur in the region $\rho>1$. It is straightforward to establish the following conclusions (Figure \ref{fig:2}):

\begin{proposition}\label{Prop-SR}
\bi 
\item[(i)] For $b\in[0,1]$, $\lim_{\rho\to 1^+}l(\rho,b)=\lim_{\rho\to +\infty}l(\rho,b)=+\infty$ and $l(\rho,b)$ has a unique local minimum in $(1,+\infty)$ at 
\be \rho=\rho_+(b) = \left\{ 
\begin{array}{cc}
    \frac{\sqrt{b^2-10b+13}-(1+b)}{2(1-b)}, & b\in[0,1); \\
    3/2, & b=1, 
\end{array} \right.
\label{rplus}
\ee
whereat $\lambda_{\rm{min}}=l(\rho_+,b)\in(6.75,26.87)$. Thus the \godel\ profile applies in this case, with $n_0\in\{3,4,5,6\}$ depending on the value of $b$. 
\item[(ii)] For $b\in (1,5-\sqrt{12})$, 
$\lim_{\rho\to 1^+}l(\rho,b)=+\infty$ and $\lim_{\rho\to +\infty}l(\rho,b)=0$. $l(\rho,b)$ has a unique local minimum and a unique local maximum in $(1,+\infty)$. Thus for each $n\geq 1$, there is at least one value of $r>\alpha^{-1}$ for which $\lambda(r)=n^2$ (and all spin-vectors on the corresponding circular CTC are $T_\gamma-$periodic). For a finite subset of positive integers (which may be empty), there are three solutions of $\lambda(r)=n^2$ with $r>\alpha^{-1}$ (and three corresponding circular CTCs with all spin-vectors periodic); for all other values of $n$, there is exactly one solution and one corresponding CTC. 
\item[(iii)] For $b\in[5-\sqrt{12},2)$, $l(\rho,b)$ is a decreasing function of $\rho$, with $\lim_{\rho\to 1^+}l(\rho,b)=+\infty$ and $\lim_{\rho\to +\infty}l(\rho,b)=0$. Hence for each positive integer $n\geq1$, there is a unique solution of $\lambda(r)=n^2$ with $r>\alpha^{-1}$, and correspondingly, a unique circular CTC along which all spin-vectors are $T_\gamma-$periodic. 
\ei
\hfill$\blacksquare$
\end{proposition}

\subsubsection{Van Stockum's Solution and Tipler machines}

Van Stockum's solution \cite{van1937gravitational}, representing an infinitely long rigidly rotating cylinder of, is the uncharged ($\beta=0$) Som-Raychaudhuri solution. So (\ref{SR-lambda}) holds, and Proposition \ref{Prop-SR} holds with $b=0$. Thus the \godel\ profile applies, and we find $n_0=6$.





Van Stockum showed how his line element can be matched to that of a vacuum spacetime, and Tipler subsequently demonstrated that this construction gives rise to cylindrical structures of finite radius that operate as time machines \cite{tipler1974rotating}. The exterior line element again has the form (\ref{stn-cyl-lel}), but with 
\begin{eqnarray} 
F(r)&=&\frac{r\sin(\beta-\mu)}{R\sin\beta},\label{F-VS}\\
H_1(r)=H_2(r) &=& e^{-\alpha^2R^2}\left(\frac{r}{R}\right)^{-2\alpha^2R^2},\label{H-VS}\\
L(r) &=& \frac{rR\sin(3\beta+\mu)}{2\sin2\beta\cos\beta},\label{L-VS}\\
M(r)&=& \frac{r\sin(\beta+\mu)}{\sin2\beta},\label{M-VS}
\end{eqnarray}
where $\alpha R>1/2$ and 
\begin{eqnarray} \mu(r)&=&(4\alpha^2R^2-1)^{1/2}\ln\left(\frac{r}{R}\right),\label{VS-gamma} \\
\beta &=& \arctan(4\alpha^2R^2-1)^{1/2}. \label{VS-beta} 
\end{eqnarray}
Note that since $\alpha>0$ and $\beta\in(0,\pi/2)$, we can write 
\be \sin\beta = \frac{\sqrt{4\alpha^2R^2-1}}{2\alpha R},\quad \cos\beta=\frac{1}{2\alpha R} \label{trigs}\ee
and so 
\be 2\sin2\beta\cos\beta = \frac{\sqrt{4\alpha^2R^2-1}}{2\alpha^3R^3}.\label{trigs2} \ee
Comparing (\ref{L-VS}), we see that circular CTCs (which occur if and only if $L(r)<0$) occur if and only if $\sin(3\beta+\mu)<0$. This oscillatory behaviour means that as $r$ increases, we move through regions that are alternately filled with and devoid of circular CTCs. 

The spacetime with line element implied by (\ref{F-VS})-(\ref{VS-beta}) matches smoothly across the hypersurface $\{r=R\}$ with the van Stockum spacetime. 

We calculate 

\be \lambda(r) = -\frac12e^{\alpha^2r^2}\left(\frac{r}{R}\right)^{2\alpha^2R^2-1}\alpha^3R^3\sqrt{4\alpha^2R^2-1}\csc(3\beta+\mu). \label{VSE-lambda} \ee
It is convenient to introduce $\rho=\alpha r$, $\ro=\alpha R$ and $\lamb(\rho)=\lambda(r)$. Then $\ro>1/2$, and we are concerned only with $\rho>\ro$, the vacuum region outside the van Stockum cylindrical region. We can establish the following.

\begin{proposition}\label{Prop-VS}
\bi
\item[(i)] For $\ro\in(1/2,1]$, $\lamb(\rho)$ is positive on the sequence of intervals $\{I_m\}_{m=1}^\infty$, where 
\be I_m = \left(\ro\exp\left(\frac{(2m-1)\pi-3\sqrt{4\ro^2-1}}{\sqrt{4\ro^2-1}}\right),\ro\exp\left(\frac{2m\pi-3\sqrt{4\ro^2-1}}{\sqrt{4\ro^2-1}}\right)\right).\label{Im-def}
\ee
On each interval $I_m$, $\lamb(\rho)$ approaches $+\infty$ at each endpoint, and has a unique minimum in $I_m$. It follows that the \godel\ profile obtains on each interval, with the value of $n_0$ depending on both $\ro$ and $m$. 
\item[(ii)] For $\ro>1$, we have the same conclusion as in part (i), but with $I_1$ given by 
\be I_1 = \left(\ro, \ro\exp\left(\frac{2\pi-3\sqrt{4\ro^2-1}}{\sqrt{4\ro^2-1}}\right)\right). \label{I1-def} 
\ee
\ei
\hfill$\blacksquare$
\end{proposition}

The essential difference between these two cases is that for $\ro>1$, there are circular CTCs that are arbitrarily close to the matter filled region $\rho\leq\ro$. For the lower values of $\ro$, we must move farther into the vacumm region before meeting these CTCs. 

Recall that the parameter $n_0$ of the \godel\ profile counts how many positive integers do not contribute a pair of circular CTCs along which all spin-vectors are periodic. In the present example, the value of $n_0$ grows rapidly with $m$: for example, with $\ro=1$, we find 
\begin{eqnarray} \lambda_{\rm{min}}|_{I_1} &\simeq& 13.70, \\
\lambda_{\rm{min}}|_{I_2} &\simeq& 8\times 10^{619}, \\
\lambda_{\rm{min}}|_{I_3} &\simeq& 10^{870,109}.
\end{eqnarray}
The corresponding parameter $n_0$ is given by the integer ceiling of the square root of these values. 

\subsection{Kerr spacetime}
In Boyer-Lindquist coordinates $x^a=(\tau,r,\theta,\phi)$, the line element of Kerr spacetime reads
\be ds^2 =\rho^2\left(\frac{dr^2}{\Delta}+d\theta^2\right)+(r^2+a^2)\sin^2\theta d\phi^2 - d\tau^2 + \frac{2Mr}{\rho^2}(a\sin^2\theta d\phi-d\tau)^2, \label{eq:kerr-lel} \ee
where
\be \rho^2=r^2+a^2\cos^2\theta,\quad \Delta = r^2-2Mr+a^2. \label{eq:kerr-metric-fns} \ee
As shown by Carter \cite{carter1968global}, the region $r<r_-=M-\sqrt{M^2-a^2}$ is totally vicious. We impose the black hole condition $M\geq|a|$ and we focus on a simple class of CTCs: those for which $\tau,r$ and $\theta$ are constant along the curve (we'll maintain the name circular CTCs for these curves, as well as the label $\gamma$). A necessary and sufficient condition for these closed curves to be timelike is that 
\be g_{\phi\phi} = (r^2+a^2)\sin^2\theta+\frac{2Ma^2r}{\rho^2}\sin^4\theta<0. \label{g-phi-phi} 
\ee
Notice that this requires $r<0$. We write $g_{\phi\phi}=-\beta^2, \beta>0$, and so the tangent to the circular CTC is $u^a = \beta^{-1}\left(\pd{}{\phi}\right)^a$, which has proper-time period $T_\gamma=2\pi\beta$.

We can summarise the properties of solutions of the Fermi-Walker transport equations along $\gamma$ as follows. Here, $s^i, i=0,1,2,3$ refers to the Boyer-Lindquist coordinate components of the spin-vector $s^a$. In this description, \textit{constants} refer to quantities that depend on the fixed values of $r$ and $\theta$ along $\gamma$,  and on the black hole parameters $M,a$. 
\bi
\item[(i)] The orthogonality condition $g_{ab}u^as^b=0$ shows that $s^0$ is a constant multiple of $s^3$. 
\item[(ii)] There exists a solution of the FW transport equations (unique up to trivial scalings) along which $s^0=s^3=0$ and $s^1,s^2$ are constants. This yields a periodic spin-vector, as guaranteed by Proposition \ref{prop2}.
\item[(iii)] The remaining two linearly independent solutions of the FW transport equations are determined by solutions of the equation
\be S''(t)+\lambda^2 S(t) =0,\label{master:kerr} \ee
where
 
 \be \lambda^2 = 
\frac{4 a^2 M^2 \sin ^2\theta\left(A+B+C\right)}{\left(a^2 \cos 2 \theta +a^2+2 r^2\right)^3 \left(a^4+a^2 \cos 2\theta \left(a^2+r (r-2 M)\right)+a^2 r (2 M+3 r)+2 r^4\right)^2}, \label{lambda:kerrs}
 \ee
 with 
\begin{eqnarray}
A &=& (3 a^8-10 a^6 r^2-a^4 r^3 (8 M+r)+72 a^2 r^6+72r^8,\label{Adef}\\
B&=&a^4 \cos 4 \theta  \left(a^4-6 a^2 r^2+r^3 (8 M-3 r)\right), \label{Bdef}\\
C&=& 4 a^2 \left(a^6-4 a^4 r^2-3 a^2 r^4+6 r^6\right) \cos
   2\theta. \label{Cdef}
\end{eqnarray}
\ei

As in the introductory paragraph of Subsection III-A, we can then phrase (spacetime) periodicity of the spin vector in terms of an equation that requires an integer value for the ratio of the period of the CTC and the period of the spin-vector. This equation takes the form $-\lambda^2g_{\phi\phi}=n^2, n\in\mathbb{N}$. If we specialise to the equatorial plane, we obtain the following:  

\begin{proposition}\label{prop:kerr} Let $\gamma$ be a circular CTC of Kerr spacetime lying in the equatorial plane, so that $\theta=\frac{\pi}{2}$ along $\gamma$. Then $r\in(r_*,0)$, where $r_*$ is the unique negative root of $r^3+a^2r+2Ma^2$. Every spin-vector carried by $\gamma$ is $T_\gamma$-periodic if and only if 
\be \Gamma(r)= -\frac{a^2M^2(a^2+3r^2)^2}{r^5(r^3+a^2r+2Ma^2)}=n^2 \quad\hbox{ for some } n\in\mathbb{N}. \label{kerr-key}
\ee
If this equation has no solutions, then there is a unique spin-vector that is $T_\gamma$-periodic along $\gamma$. Furthermore the \godel\ profile applies: $\Gamma(r)\to+\infty$ as $r\to r_*^+$ and as $r\to 0^-$, and has a unique positive minimum on $(r_*,0)$. \hfill$\blacksquare$ 
\end{proposition}

Since we are in the equatorial plane, the timelike condition simplifies considerably:
\be g_{\phi\phi}|_{\theta=\frac{\pi}{2}} = r^2+a^2 +\frac{2Ma^2}{r}<0. 
\label{eq:ctc-equator}
\ee
Thus as claimed above, circular CTCs arise on the equatorial plane for $r\in(r_*,0)$ where $r_*$ is the unique negative root of $r^3+a^2r+2Ma^2$. Note that this last expression is positive on $(r_*,0)$, and so $\Gamma(r)$ is positive on this interval. Clearly, $\Gamma(r)\to+\infty$ as $r\to r_*^+$ and as $r\to 0^-$. Less clear, but verifiable algebraically, is the fact that $\Gamma$ has a unique minimum in $(r_*,0)$ for all $M>|a|>0$. This implies that the \godel\ profile also applies in this case. Numerical experimentation across the $(M,a)$ parameter space indicates a minimum value of $n_0$ of 5 which applies in the extremal limit $|a|=M$, with $n_0$ unbounded in the limit $|a|\to0$. The results summarised here are established by writing $r=uM, |a|=vM$, in which case 
\be \Gamma = -\frac{v^2(3u^2+v^2)^2}{u^5(u^3+v^2u+2v^2)},\quad v\in (0,1], \quad u\in (u_*,0), \label{rewrite} 
\ee
where $u_*$ is the unique negative root of $u^3+v^2u+2v^2$. Applying an asymptotic balance argument yields 
\be \Gamma_{min} \sim 3\left(\frac{2}{v}\right)^{4/3},\quad v\to 0. \label{eq:small-a} 
\ee
This limit corresponds to $a\to0$ at fixed $M$. As above, the integer ceiling of the square root of this term gives the value of $n_0$ for the \godel\ profile. 

\subsection{Taub-NUT spacetime}

Taub-NUT spacetime provides an interesting contrast to the behaviour seen in the examples reviewed so far. Following \cite{hawking1973large}, we consider the spacetime $\st=\mathbb{R}\times\mathbb{S}^3$ with line element 
\be ds^2 = -U^{-1}d\tau^2+4\ell^2U(d\psi+\cos\theta d\phi)^2+(t^2+\ell^2)(d\theta^2+\sin^2\theta d\phi^2), \label{eq:TNUT} \ee
where 
\be U(\tau)  = -\left(\frac{\tau^2-2M\tau-\ell^2}{\tau^2+\ell^2}\right), \quad \tau\in\mathbb{R} \label{eq:udef} \ee 
and $(\theta,\phi,\psi)$ are Euler angles on $\mathbb{S}^3$. Note in particular that $\psi\in[0,4\pi]$ with $4\pi-$periodic identification. See \cite{griffiths2009exact} for a review of the global properties of this spacetime which discusses (among other things) the respective roles of the mass parameter $M$ and the NUT parameter $\ell$. 

This spacetime provides a counterexample to almost anything \cite{misner1967taub}, and chronology violation is no exception. Circular CTCs, along which $\tau,\theta$ and $\phi$ are constant, are present in regions where $U<0$. This corresponds to the regions 
\be \tau<\tau_-= M-\sqrt{M^2+\ell^2}<0,\quad \tau>\tau_+=M+\sqrt{M^2+\ell^2}>0. 
\label{TNUT-regions} 
\ee
These CTCs have proper-time period $T_\gamma=8\pi\ell\sqrt{-U}$.  

In coordinates $(x^0,x^1,x^2,x^3)=(\tau,\psi,\theta,\phi)$, the solutions of the FW transport equations have these properties: 

\bi
\item[(i)] $s^1$ is a constant multiple of $s^3$.
\item[(ii)] $s^0$ is constant along all solutions, and the unique solution with  $s^1=s^2=s^3=0$ provides a periodic spin-vector as guaranteed by Proposition \ref{prop2}.
\item[(iii)]The remaining two linearly independent solutions of the FW transport equations are determined by solutions of the equation
\be S''(t)+\frac{\ell^2(\tau^2-2M\tau-\ell^2)}{(\ell^2+\tau^2)^3} S(t) =0.\label{master:tnut} \ee
\ei

Differences relative to Kerr spacetime arise when we look at the existence condition for spacetime periodic spin-vectors analogous to (\ref{kerr-key}). In the Taub-NUT case, this reads 
\be 4\ell^2\left(\frac{\tau^2-2M\tau-\ell^2}{(\tau^2+\ell^2)^2}\right) = n\in \mathbb{N}, 
\label{TNUT-key}
\ee
or equivalently (with $\tau=uM, \ell = vM$) 
\be n = G(u,v) = 4v^2\left(\frac{u^2-2u-v^2}{(u^2+v^2)^2}\right),\quad n\in\mathbb{N},\quad v\in\real,\quad u\in(-\infty,u_-)\cup(u_+,+\infty), 
\label{G-TNUT} 
\ee
where $u_\pm=1\pm\sqrt{1+v^2}$. Some elementary calculus yields the following results. 

\begin{proposition} \label{prop:tnut}
\bi
\item[(i)] For $u>u_+$, $G(u,v)<\frac12$ for all $v\in\mathbb{R}$. Hence there are no solutions of (\ref{TNUT-key}) for $\tau>\tau_+$, and so there is a single $T_\gamma$-periodic spin-vector along any circular CTC $\gamma$ in this region. 
\item[(ii)] For each $v\in\real$, $G(u,v)$ has a unique positive maximum on $(-\infty,u_-)$ and vanishes in the limit at the endpoints of the interval. Hence there is at most a finite number of values of $n$ which yield solutions of (\ref{G-TNUT}) for $u$, and so only a finite number of circular CTCs along which all spin-vectors are $T_\gamma$-periodic.
\ei
\hfill$\blacksquare$
\end{proposition}

\begin{figure}[ht!]
\begin{center}
{\includegraphics[width=8cm]{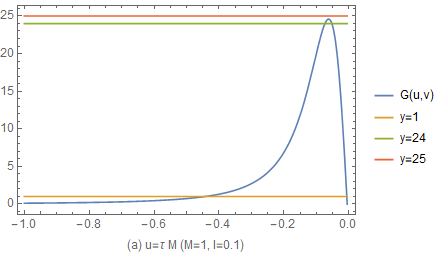},
\includegraphics[width=8cm]{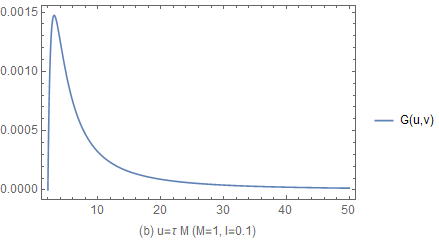}}
{\includegraphics[width=8cm]{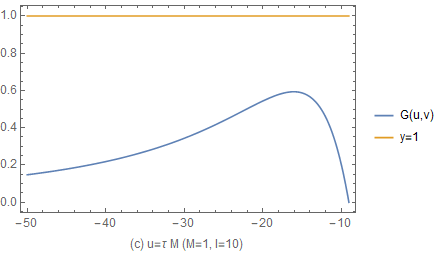},
\includegraphics[width=8cm]{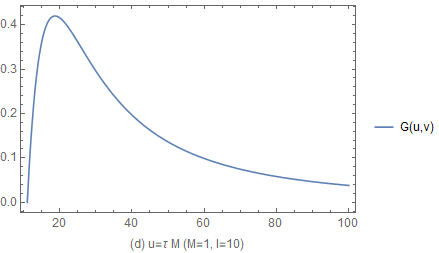}}
  \end{center}
\caption{Plots of the function $G(u,v)$ of Proposition 
\ref{prop:tnut} for Taub-NUT spacetime. Intersections of the graph with horizontal lines $y=n, n\in\mathbb{N}$ correspond to the circular CTCs along which all spin-vectors are $T_\gamma$-periodic. In all cases, we have taken $M=1$. Panels (a) and (b) show $G$ for a small value of the NUT parameter, $\ell=0.1$. As stated in Proposition \ref{prop:tnut}, there are no intersections for $\tau>\tau_+$, and a finite number ($48$ in this case) for $\tau<\tau_-$. Panels (c) and (d) show the corresponding graphs for a large value, $\ell=10$, of the NUT parameter. In this case, there are no intersections, and so for these parameter values $(M,\ell)=(1,10)$, there are no circular CTCs for which gyroscopic motion is generically consistent.}
\label{fig:3}
\end{figure}

In the limit of small $v$ (small NUT parameter $\ell$ for $M$ fixed), the maximum of $G$ has the asymptotic behaviour 
\be G_{max} \sim \frac{3\sqrt{3}}{2v},\quad v\to 0 \label{Gmax-limit}
\ee
and so the number of solutions that arises in part (ii) of Proposition \ref{prop:tnut} is unbounded in this limit. On the other hand, in the limit of large $v$, we find 
\be G_{max} \sim \frac12+\frac{\sqrt{3}}{2v},\quad y\to+\infty,
\label{Gmax-large-limit} 
\ee
and so the number of solutions is zero when $v$ is sufficiently large. See Figure \ref{fig:3}. So in this example, there are only finitely many circular CTCs along which gyroscopic motion is generically consistent. 

\subsection{Ori's asymptotically flat time-machine}

We'll mention one last example: Ori's asymptotically flat time machine \cite{ori2007formation}. This spacetime avoids many features that render the examples above unrealistic in various ways. The spacetime evolves from regular initial data. It possesses a vacuum core, a dust-filled intermediate region (obeying both the weak and strong energy conditions) and an asymptotically flat vacuum exterior. The CTCs evolve in the spacetime: none are present at early stages of the evolution from the initial data surface. They lie to the future of a chronology horizon that appears not to be compactly generated, and so avoids an associated instability \cite{hawking1992chronology}. The CTCs reside in the inner vacuum region, whose line element is given in coordinates $x^a=(v,r, \theta,\phi)$ by 
\be ds^2 = \left(1-\frac{2\mu}{r}\right)dv^2+2dvdr+r^2\left(d\theta^2+\sinh^2\theta d\phi^2\right). \label{eq:ori} 
\ee
The null coordinate $v$ satisfies $0\leq v <\ell$, with $\ell-$periodic identification. The parameter $\ell$ relates to the matter content of the dust-filled region, and has a positive lower bound, $\ell_{min}$: there is an interval's worth of choices for the value of $\ell$. The coordinate $\phi$ is the usual azimuthal coordinate, and $\theta\geq 0$.  The parameter $\mu$ is positive, and we note that $r$ is a time coordinate for $r>2\mu$. 

There are CTCs $\gamma$ in the region $r<2\mu$ along which $r,\theta$ and $\phi$ are constant. A straightforward calculation shows that in the coordinates of (\ref{eq:ori}), the components of a spin-vector along such a CTC are constant. Hence \textit{all} spin-vectors are $T_\gamma$-periodic for this family of CTCs. However, this feature disappears if we adjust the CTC in the following way. The 2-surfaces of constant $r$ and $\theta$ are topological torii. We can consider the CTCs above to traverse toroidal circles (going the long way around the doughnut). We can also construct CTCs by allowing these curves to also traverse the poloidal circles of the torus (sprialling in towards the doughnut hole and out again, while simultaneously completing a circuit the long way around). In the coordinates of (\ref{eq:ori}), these curves have tangent 
\be u^a = (\alpha,0,0,\beta), \label{eq:ori-ctc-other} \ee
with the constants $\alpha, \beta$ subject to the (unit) timelike condition
\be 
\left(1-\frac{2\mu}{r}\right)\alpha^2+r^2\sinh^2\theta \beta^2 = -1 \label{ori-ctc1} \ee
and the condition 
\be \hbox{there exist}\quad  n_1,n_2\in\mathbb{N}\quad \hbox{ such that}\quad 2\pi n_1\alpha=n_2\ell\beta, \label{ori-ctc2} \ee
which expresses the need for the periods of the toroidal and poloidal loops to be synchronised (the particle completes $n_1$ circuits of the toroidal circle and at the same time completes $n_2$ circuits of the poloidal circle).

The FW transport equations admit a constant solution (providing a spacetime periodic spin-vector along the CTC), and two independent solutions with period $T_s=\frac{2\pi}{\lambda}$, where
\be \lambda^2 = \frac{\alpha ^2 \text{csch}^2\theta \left(\left(\alpha ^2-1\right) r-2 \alpha ^2 \mu \right) \left(9 \mu ^2+2 r^2+\mu  \cosh 2 \theta (4 r-9 \mu )-8 \mu  r\right)}{2 r^5}.
\label{l2:ori}
\ee

Since the proper-time period of the CTC is $T_\gamma = n_1\ell/\alpha = 2\pi n_2/\beta$, we see that the spacetime periodicity condition for the spin vector, $T_\gamma = n T_s$, brings us back to the familiar territory of a countable (possible finite, possibly empty) set of solutions for $n$, for a given set of values of the parameters defining the CTC (the constants $r,\theta,\mu,\ell,\alpha,\beta,n_1,n_2$, satisfying (\ref{ori-ctc1}) and (\ref{ori-ctc2})). 

\section{Conclusions}

In relation to the examples studied above, it is clear that periodicity (and hence consistency) of gyroscopic motion is a very special property. Among the families of CTCs considered, indexed by a set of continuous real parameters, there are only (at most) a countable number of CTCs for which gyroscopic motion is consistent. Ori's spacetime considered in section III.D is an exception. But a small perturbation shows a return to the outcome just described. So for these examples, we can say that in a mathematically precise sense, gyroscopic motion is generically inconsistent: on an open, dense subset of the 3-parameter families of CTCs, there is an open dense subset of the the initial data set (which recall is $\mathbb{S}^2$) for gyroscopic motion for which the corresponding solutions of the equations of motion do not return to their initial position on completion of a circuit of the CTC. This is our consistency requirement, and it is generically not met. 

But we would like to be able to say more. After all, the examples above are just that - examples - and may not represent the generic behaviour. We consider evidence for the following conjecture: \textit{the motion of gyroscopes along CTCs in a time-orientable spacetime is generically inconsistent}. 

Validating this conjecture would require the statement and proof of a result that demonstrates the exceptional nature of a CTC along which gyroscopic motion is consistent. It seems clear that an important necessary aspect of such a statement is valid in the following sense. Suppose that a region of spacetime is filled with a congruence of CTCs, and let $\vec{\alpha}\in O$ parametrise the different members of the congruence (with $O$ an open subset of $\mathbb{R}^3$). (All the examples above have this feature.) By standard ODE theory, solutions of the FW transport equations depend continuously on the parameters $\vec{\alpha}$ (see e.g.\ section 2.3 of \cite{perkodifferential}). It follows that the eigenvalues $\{1,e^{\pm i\theta}\}$ of the monodromy matrix $M^{a'}_{b'}$ depend continuously on the parameters $\vec{\alpha}$ (cf.\ the discussion preceding Proposition \ref{prop2} above). So if a CTC with parameters $\vec{\alpha}_0$ has generically inconsistent gyroscopes ($\theta(\vec{\alpha}_0)\neq0$), then there is an open neighbourhood $O'\subset O$ of $\vec{\alpha}_0$ for which all CTCs with parameters $\vec{\alpha}\in O'$ also have generically inconsistent gyroscopes. Thus CTCs with generically inconsistent gyroscopes occur in open sets. We would also like to show that the set of CTCs with generically inconsistent gyroscopes are dense in the set of all CTCs, which would require a thorough understand of the structure of these sets. A way to approach this would be to show that a generic perturbation of a CTC with generically consistent gyroscopes leads to one with generically inconsistent gyroscopes. This appears to be the case in the example of Ori's spacetime: purely toroidal CTCs have consistent gyroscopes, but a perturbation that generates poloidal motion of the CTC yields generically inconsistent gyroscopic motion. This example also appears to show that there is some subtlety in the problem: the `right' perturbation must be identified - but presumably generic perturbations would include these. Combining the geometric approach to perturbations in spacetime (that is, of spacetimes themselves and lower dimensional submanifolds of those spacetimes) with the general theory of perturbations of linear operators \cite{kato2013perturbation} may provide the means to analyse this situation more thoroughly - recall that the FW transport equations are linear. 

Does it matter that a gyroscope carried by a CTC may not be consistently defined? Such an inconsistency may generate what we could call `practical' difficulties for an associated time traveller: their compass of inertia will rotate around a fixed direction as they depart from and return to a spacetime event $P$. So while their notion of `up-down' may remain the same, those of `left-right' and `forwards-backwards' will be altered, despite their being at the same spacetime event. For external observers of the gyroscope, there are what could be considered more fundamental difficulties relating to paradoxes of identity induced by time travel (\cite{wasserman2017paradoxes}, p.\ 21). Let us imagine an observer and a gyroscope crossing a chronological horizon. Imagine further that the observer continues on a causally well-behaved path (an open timelike curve), but the gyroscope moves onto a CTC. The observer will immediately see the gyroscope pointing in numerous different directions, as many as the number of loops of the CTC it traverses. While this outcome is a more mundane consequence of time travel than (say) the situation envisaged in the grandfather paradox, it appears to be harder to avoid: the motion of the gyroscope is simply and directly governed by physical laws in the form of the FW transport equation. It suggests that when taking extended body effects into consideration, consistency in the presence of CTCs might not be possible.

\acknowledgements
Thanks to Abraham Harte, Peter Taylor, Ko Sanders and Erik Curiel  for useful conversations.

\appendix

\section{Proof of the Geometric Floquet-Lyapunov Theorem}

Let $\fab{a}{b}$ be as in the statement of the proposition, and let $\eab{a}{b}$ be the exponential matrix solving the IVP 
\be \frac{dF^{a'}_{b'}}{dt} = B^{a'}_{c'}F^{c'}_{b'},\quad F^{a'}_{b'}(0) =\delta^{a'}_{b'},\label{exp-F}\ee
where $B^{a'}_{b'}$ is a non-singular type (1,1) tensor at $z$. We call a type (1,1) tensor $B^{a'}_{b'}$ non-singular if there exists another type (1,1) tensor $\tilde{B}^{a'}_{b'}$ such that 
\be B^{a'}_{c'}\tilde{B}^{c'}_{b'}
=B^{c'}_{b'}\tilde{B}^{a'}_{c'}=\delta^{a'}_{b'}. \label{non-singular} \ee

Appealing to uniqueness of solutions of linear ODEs, we see that 
\be \fab{a}{b}(t+T) = \fab{a}{c}(t)\fab{c'}{b}(T) \label{f-add-rule}\ee
for all $t\in\real$ and 
\begin{eqnarray} \eab{a}{b}(t_1+t_2)&=&\eab{a}{c}(t_1)\eab{c}{b}(t_2) \nonumber \\
&=& \eab{a}{c}(t_2)\eab{c}{b}(t_1) 
\label{e-add-rule}
\end{eqnarray}
for all $t_1,t_2 \in\mathbb{R}$. 
Now define 
\be \gtab{a}{b}(t) = \fab{a}{c}(t)\eab{c}{b}(t). \ee
Then using (\ref{f-add-rule}) and (\ref{e-add-rule}) we have 
\be
\gtab{a}{b}(t+T) = \fab{a}{d}(t)\fab{d'}{c}(T)\eab{c}{e}(T)\eab{e}{b}(t) 
\label{gt-add-rule}
\ee
which equates to $\gtab{a}{b}(t)$ - establishing $T-$periodicity of this quantity - provided 
\be \fab{d'}{c}(T)\eab{c}{e}(T) = \delta^{d'}_{e'}. \label{delta-con} \ee

We show that there exists a non-singular type (1,1) tensor $B^{a'}_{b'}$ so that this is satisfied. 

This can be done by introducing a pseudo-orthonormal basis of $T_z(\st)$, $\{e^{a'}_{(i)}, i=1,2,3,4\}$, so that 
\be g_{a'b'}(z)e^{a'}_{i}e^{b'}_j = \eta_{ij}, \label{poob} \ee
where $\eta$ is the Minkowski unit tensor. We also introduce the corresponding basis of 1-forms
\be \omega^i_{a'} = g_{a'b'}e^{b'}_i. \label{one-forms} \ee
Projecting onto this basis, we can write (\ref{exp-def-de}) as a matrix-valued scalar equation
\be F'(t) = B F(t),\quad F(0)=I_4 \label{exp-prop-matrix} \ee
where $F = (F^i_j)$ is a $4\times 4$ matrix with components 
\be F^i_j = F^{a'}_{b'}\omega^i_{a'}e^{b'}_j,\ee
and likewise for $A$, which is a \textit{constant} matrix. $I_4$ is the $4\times 4$ identity matrix. Then we can write down the solution of (\ref{exp-prop-matrix}):
\be F(t) = \exp(tB). \label{E-matrix-sol} \ee
Let $G=(G^i_j)$ be the projection of $\fab{a}{b}(T)$ on the pseudo-orthonormal basis. Then (\ref{delta-con}) reads 
\be G(T)F(T)=G(T)\exp(TB)=I_4. \label{matrix-delta-con} \ee
Since the matrix $G(T)$ is non-singular (by virtue of (\ref{Liouville})), it follows that there is a non-singular matrix $B$ satisfying this equation (see e.g.\ Section 3.1 of \cite{coddington1955theory}). With this choice of $B$, we see that 
\be \gtab{a}{b}(t+T) = \gtab{a}{b}(t). \label{gt-periodic} \ee
Now define 
\be E^{a'}_{b'}(t) = \eab{a}{b}(-t),\quad t\in\mathbb{R}. \label{E-exp-def} \ee
It is straightforward to show that this is an exponential matrix, satisfying (\ref{exp-def-de}) with $A^{a'}_{b'}=-B^{a'}_{b'}$. Using (\ref{e-add-rule}),  we can write 
\be \fab{a}{b}(t) = \gtab{a}{c}(t)E^{c'}_{b'}(t), \label{final}
\ee
which completes the proof. \hfill$\blacksquare$

\section*{References}
\bibliographystyle{unsrt}
\bibliography{mybib}

\begin{thebibliography}{10}

\bibitem{friedman1990cauchy}
John Friedman, Michael~S Morris, Igor~D Novikov, Fernando Echeverria, Gunnar
  Klinkhammer, Kip~S Thorne, and Ulvi Yurtsever.
\newblock {Cauchy problem in spacetimes with closed timelike curves}.
\newblock {\em Physical Review D}, 42(6):1915, 1990.

\bibitem{echeverria1991billiard}
Fernando Echeverria, Gunnar Klinkhammer, and Kip~S Thorne.
\newblock {Billiard balls in wormhole spacetimes with closed timelike curves:
  Classical theory}.
\newblock {\em Physical Review D}, 44(4):1077, 1991.

\bibitem{novikov1992time}
Igor~D Novikov.
\newblock {Time machine and self-consistent evolution in problems with
  self-interaction}.
\newblock {\em Physical Review D}, 45(6):1989, 1992.

\bibitem{levanony2011extended}
Dana Levanony and Amos Ori.
\newblock {Extended time-travelling objects in Misner space}.
\newblock {\em Physical Review D}, 83(4):044043, 2011.

\bibitem{lewis1976paradoxes}
David Lewis.
\newblock {The paradoxes of time travel}.
\newblock {\em American Philosophical Quarterly}, 13(2):145--152, 1976.

\bibitem{harte2015motion}
Abraham~I Harte.
\newblock {Motion in classical field theories and the foundations of the
  self-force problem}.
\newblock In {\em Equations of Motion in Relativistic Gravity}, pages 327--398.
  Springer, 2015.

\bibitem{dixon1970dynamics1}
William~G Dixon.
\newblock {Dynamics of extended bodies in general relativity. I. Momentum and
  angular momentum}.
\newblock {\em Proceedings of the Royal Society of London. A. Mathematical and
  Physical Sciences}, 314(1519):499--527, 1970.

\bibitem{dixon1970dynamics2}
William~G Dixon.
\newblock {Dynamics of extended bodies in general relativity-ii. moments of the
  charge-current vector}.
\newblock {\em Proceedings of the Royal Society of London. A. Mathematical and
  Physical Sciences}, 319(1539):509--547, 1970.

\bibitem{dixon1974dynamics3}
William~G Dixon.
\newblock {Dynamics of extended bodies in general relativity III. Equations of
  motion}.
\newblock {\em Philosophical Transactions of the Royal Society of London.
  Series A, Mathematical and Physical Sciences}, 277(1264):59--119, 1974.

\bibitem{misner1973gravitation}
Charles~W Misner, Kip~S Thorne, and John~Archibald Wheeler.
\newblock {\em {Gravitation}}.
\newblock Macmillan, 1973.

\bibitem{misner1967taub}
Charles~W Misner.
\newblock Taub-nut space as a counterexample to almost anything.
\newblock {\em Relativity theory and astrophysics}, 1:160, 1967.

\bibitem{wald1984general}
RM~Wald.
\newblock {\em General relativity}.
\newblock Chicago, University of Chicago Press, 1984.

\bibitem{minguzzi2019lorentzian}
Ettore Minguzzi.
\newblock {Lorentzian causality theory}.
\newblock {\em Living Reviews in Relativity}, 22(1):3, 2019.

\bibitem{wasserman2017paradoxes}
Ryan Wasserman.
\newblock {\em {Paradoxes of time travel}}.
\newblock Oxford University Press, 2017.

\bibitem{yakub1975}
V.A. Yakubovich and V.M. Starzhinskii.
\newblock {\em Linear differential equations with periodic coefficients}.
\newblock John Wiley and Sons, 1975.

\bibitem{magnus2013hill}
Wilhelm Magnus and Stanley Winkler.
\newblock {\em Hill's equation}.
\newblock Courier Corporation, 2013.

\bibitem{coddington1955theory}
Earl~A Coddington and Norman Levinson.
\newblock {\em Theory of ordinary differential equations}.
\newblock Tata McGraw-Hill Education, 1955.

\bibitem{godel1949example}
Kurt G{\"o}del.
\newblock {An example of a new type of cosmological solutions of Einstein's
  field equations of gravitation}.
\newblock {\em Reviews of modern physics}, 21(3):447, 1949.

\bibitem{van1937gravitational}
Willem~Jacob van Stockum.
\newblock {The gravitational field of a distribution of particles rotating
  about an axis of symmetry}.
\newblock In {\em Proc. Roy. Soc. Edinburgh}, volume~57, pages 135--154, 1937.

\bibitem{tipler1974rotating}
Frank~J Tipler.
\newblock {Rotating cylinders and the possibility of global causality
  violation}.
\newblock {\em Physical Review D}, 9(8):2203, 1974.

\bibitem{bini2019agodel}
Donato Bini, Andrea Geralico, Robert~T Jantzen, and Wolfango Plastino.
\newblock {G{\"o}del spacetime: Planar geodesics and gyroscope precession}.
\newblock {\em Physical Review D}, 100(8):084051, 2019.

\bibitem{bini2019bgodel}
Donato Bini, Andrea Geralico, Robert~T Jantzen, and Wolfango Plastino.
\newblock {G{\"o}del spacetime: elliptic-like geodesics and gyroscope
  precession}.
\newblock {\em arXiv preprint arXiv:1905.04917}, 2019.

\bibitem{griffiths2009exact}
Jerry~B Griffiths and Ji{\v{r}}{\'\i} Podolsk{\`y}.
\newblock {\em {Exact space-times in Einstein's general relativity}}.
\newblock Cambridge University Press, 2009.

\bibitem{ellis2000editor}
GFR Ellis.
\newblock {Editor's Note}.
\newblock {\em General Relativity and Gravitation}, 32(7):1399--1408, 2000.

\bibitem{ozsvath2003godel}
Istvan Ozsvath and Engelbert Schucking.
\newblock {G{\"o}del’s trip}.
\newblock {\em American Journal of Physics}, 71(8):801--805, 2003.

\bibitem{rindler2009godel}
Wolfgang Rindler.
\newblock {G{\"o}del, Einstein, Mach, Gamow, and Lanczos: G{\"o}del’s
  remarkable excursion into cosmology}.
\newblock {\em American Journal of Physics}, 77(6):498--510, 2009.

\bibitem{nolan2020causality}
Brien~C Nolan.
\newblock {Causality violation without time-travel: closed lightlike paths in
  G{\"o}del’s universe}.
\newblock {\em Classical and Quantum Gravity}, 37(8):085007, 2020.

\bibitem{som1968cylindrically}
MM~Som and AK~Raychaudhuri.
\newblock {Cylindrically symmetric charged dust distributions in rigid rotation
  in general relativity}.
\newblock {\em Proceedings of the Royal Society of London. Series A.
  Mathematical and Physical Sciences}, 304(1476):81--86, 1968.

\bibitem{paiva1987time}
FM~Paiva, MJ~Rebou{\c{c}}as, and AF~da~F Teixeira.
\newblock {Time travel in the homogeneous Som-Raychaudhuri universe}.
\newblock {\em Physics Letters A}, 126(3):168--170, 1987.

\bibitem{carter1968global}
Brandon Carter.
\newblock {Global structure of the Kerr family of gravitational fields}.
\newblock {\em Physical Review}, 174(5):1559, 1968.

\bibitem{hawking1973large}
S.W. Hawking and G.F.R. Ellis.
\newblock {\em {The large scale structure of space-time}}.
\newblock Cambridge University Press, 1973.

\bibitem{ori2007formation}
Amos Ori.
\newblock Formation of closed timelike curves in a composite vacuum/dust
  asymptotically flat spacetime.
\newblock {\em Physical Review D}, 76(4):044002, 2007.

\bibitem{hawking1992chronology}
Stephen~W Hawking.
\newblock Chronology protection conjecture.
\newblock {\em Physical Review D}, 46(2):603, 1992.

\bibitem{perkodifferential}
L.~Perko.
\newblock {\em {Differential equations and dynamical systems}}.
\newblock Springer-Verlag, New York, 1991.

\bibitem{kato2013perturbation}
Tosio Kato.
\newblock {\em Perturbation theory for linear operators}, volume 132.
\newblock Springer Science \& Business Media, 2013.

\end{thebibliography}

\end{document}